\documentclass[10pt,twocolumn,letterpaper]{article}

\usepackage[final]{iccv}      
\usepackage{makecell}  

\usepackage{multirow}

\definecolor{iccvblue}{rgb}{0.21,0.49,0.74}
\usepackage[pagebackref,breaklinks,colorlinks,allcolors=iccvblue]{hyperref}

\def\paperID{14428} 
\def\confName{ICCV}
\def\confYear{2025}

\title{Towards Scalable and Robust White Matter Lesion Localization via Multimodal Deep Learning}

\author{Julia Machnio\\
Pioneer Centre for AI\\
University of Copenhagen\\
{\tt\small juma@di.ku.dk}
\and
Sebastian Nørgaard Llambias\\
Pioneer Centre for AI\\
University of Copenhagen\\
{\tt\small snl@di.ku.dk}
\and
Mads Nielsen\\
Pioneer Centre for AI\\
University of Copenhagen\\
{\tt\small madsn@di.ku.dk}
\and
Mostafa Mehdipour Ghazi\\
Pioneer Centre for AI\\
University of Copenhagen\\
{\tt\small ghazi@di.ku.dk}
}

\begin{document}
\maketitle

\begin{abstract}

White matter hyperintensities (WMH) are radiological markers of small vessel disease and neurodegeneration, whose accurate segmentation and spatial localization are crucial for diagnosis and monitoring. While multimodal MRI offers complementary contrasts for detecting and contextualizing WM lesions, existing approaches often lack flexibility in handling missing modalities and fail to integrate anatomical localization efficiently. We propose a deep learning framework for WM lesion segmentation and localization that operates directly in native space using single- and multi-modal MRI inputs. Our study evaluates four input configurations: FLAIR-only, T1-only, concatenated FLAIR and T1, and a modality-interchangeable setup. It further introduces a multi-task model for jointly predicting lesion and anatomical region masks to estimate region-wise lesion burden. Experiments conducted on the MICCAI WMH Segmentation Challenge dataset demonstrate that multimodal input significantly improves the segmentation performance, outperforming unimodal models. While the modality-interchangeable setting trades accuracy for robustness, it enables inference in cases with missing modalities. Joint lesion-region segmentation using multi-task learning was less effective than separate models, suggesting representational conflict between tasks. Our findings highlight the utility of multimodal fusion for accurate and robust WMH analysis, and the potential of joint modeling for integrated predictions.



\end{abstract}

\section{Introduction} \label{sec:intro}

White matter hyperintensities (WMH) are pathological abnormalities of the brain's white matter that commonly present as hyperintense areas on FLAIR images and hypointensities on T1-weighted MRIs \cite{zheng2023analysis}. The total WMH burden, typically measured as lesion volume, increases with age and is recognized as a marker of early neurodegeneration. It is associated with elevated risk of Alzheimer’s disease \cite{brickman2015reconsidering}, dementia \cite{debette2010clinical}, and ischemic stroke \cite{bonkhoff2022association}, among other conditions. An accurate diagnosis requires not only the detection of the lesion, but also a detailed assessment of its volume and spatial characteristics, considered alongside the clinical context \cite{d2019white,hilal2021impact,rath2022neuroradiological,yamauchi2002significance}. Manual annotation remains the clinical gold standard; however, it is labor-intensive and not easily scalable, underscoring the need for automated tools for WMH segmentation and localization.

Multimodal MRI has become an essential tool in clinical neuroimaging by leveraging complementary information from multiple sequences, such as T1-weighted and FLAIR scans \cite{lindroth2019examining}. FLAIR images enhance lesion visibility due to cerebrospinal fluid suppression \cite{sati2012flair}, while T1-weighted scans provide superior anatomical contrast and more precise delineation of brain structures \cite{howarth2006improvement}. Although numerous studies have utilized multimodal inputs to enhance segmentation accuracy \cite{liang2021anatomical,wu2019multi}, their potential for improving lesion localization or handling missing modalities remains relatively underexplored.

Recent findings have highlighted that the spatial distribution of lesions carries significant diagnostic and prognostic value \cite{biesbroek2017lesion}. However, deriving such insights often relies on resource-intensive pipelines. For instance, Coenen et al. \cite{coenen2023strategic,coenen2024strategic} manually harmonized MRI data across cohorts to perform voxel-wise and region-of-interest analyses. While informative, these workflows are limited by their constrained scalability and reproducibility. This points to a need for automated, registration-free methods that enable accurate lesion localization directly in subject space.

Deep learning methods have demonstrated strong performance in automating WMH segmentation \cite{jiang2020deep,liang2021anatomical,wu2019multi}. Some approaches incorporate anatomical priors, such as distances to known brain landmarks \cite{ghafoorian2015small,ghafoorian2017location}, or rely on registering lesions in a common template space \cite{jiang2020deep}. In our previous work \cite{machnio2024deep}, we developed a deep learning method for segmenting anatomical regions in native space, removing the need for spatial alignment. However, that approach did not fully address multimodal integration or the trade-offs in joint lesion-region prediction for WMH localization.

In this study, we present a deep learning framework for WM lesion segmentation and localization that supports both single- and multi-modal MRI inputs. Our method performs voxel-wise segmentation of WMH and anatomical regions directly in native space. We examine four input configurations: (1) FLAIR-only, (2) T1-only, (3) concatenated FLAIR and T1, and (4) interchangeable modality training using either FLAIR or T1. Furthermore, we train unified models that jointly predict regional lesion labels, enabling direct estimation of region-wise lesion burden.

Our experiments show that while multimodal inputs improve segmentation accuracy, multi-task learning introduces a trade-off, with reduced multimodal performance compared to task-specific models. Nevertheless, multi-task models offer practical benefits, such as reduced inference time and integrated anatomical insights. Overall, our findings suggest that carefully optimized multimodal and multi-task models can provide a scalable, robust, and anatomically informed solution for WM lesion analysis in both clinical and research settings.

\section{Methods} \label{sec:methods}

\begin{figure*}[h]
\centering
\includegraphics[width=0.95\textwidth]{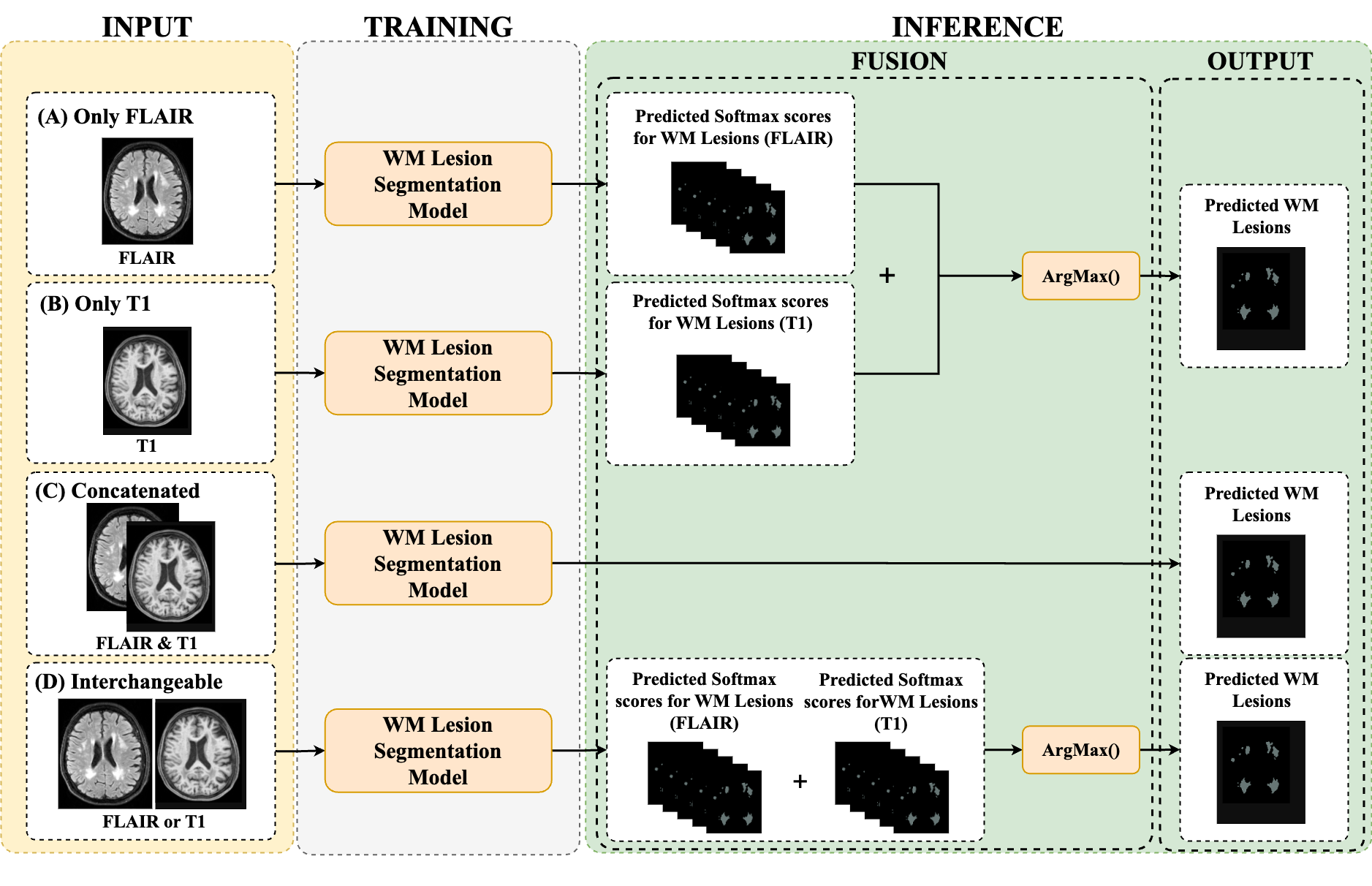}
\caption{Overview of the proposed method for WM lesion segmentation. The pipeline illustrates four input configurations used during training: (A) FLAIR only, (B) T1 only, (C) FLAIR and T1 concatenated as separate input channels, and (D) sequential training where FLAIR and T1 are treated as interchangeable modalities and passed independently through the model. For configurations (A), (B), and (D), the final prediction is obtained by voxel-wise \texttt{ArgMax} fusion across the individual softmax outputs. The same pipeline is also used for WM region segmentation.}
\label{method_diagram}
\end{figure*}

\begin{figure*}[h]
\centering
\includegraphics[width=0.95\textwidth]{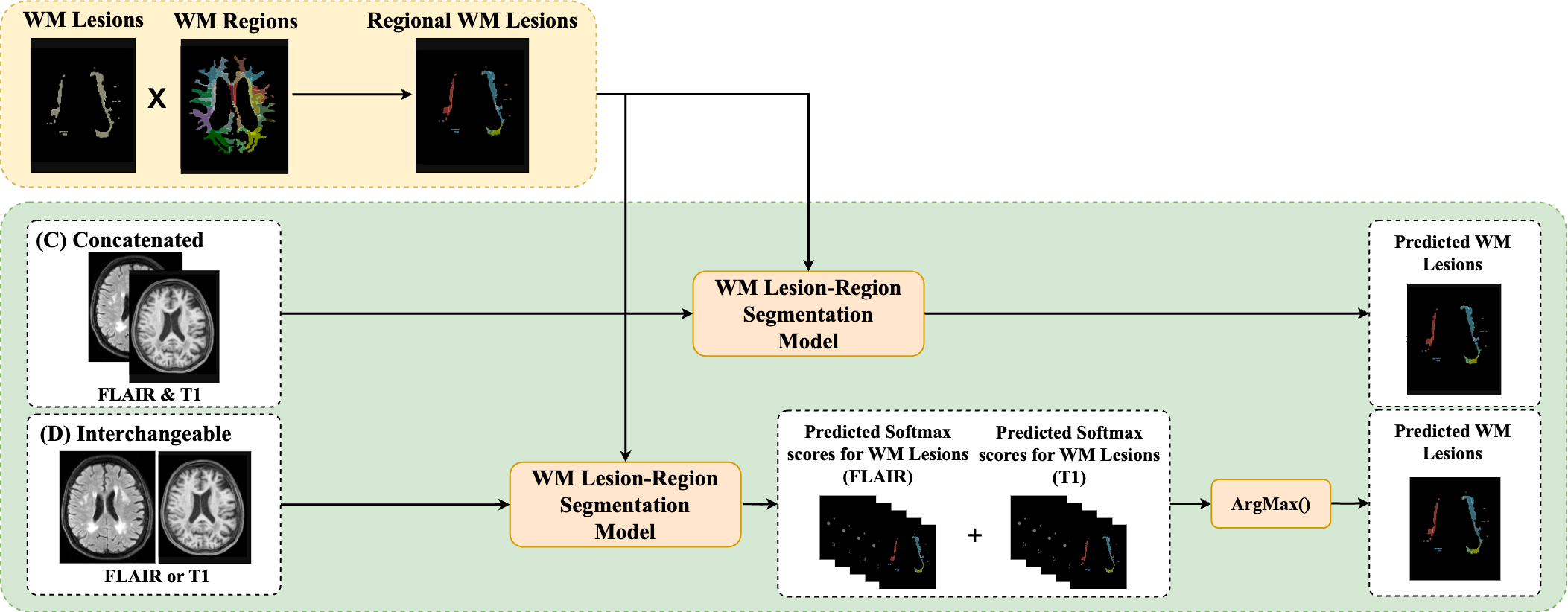}
\caption{Overview of the proposed multi-task framework for multimodal regional WMH segmentation. The pipeline adopts the same input configurations as in Figure~\ref{method_diagram} (C–D). In this setting, WM lesions and anatomical regions are jointly segmented using a unified model. Region-specific lesion labels are generated by intersecting lesion masks with WM region annotations.}
\label{method_diagram_2}
\end{figure*}

\subsection{Multimodal Configurations}

We present a deep learning approach for WM lesion segmentation and localization, utilizing both single- and multimodal MRI inputs. As illustrated in Figure~\ref{method_diagram}, we examine four input configurations: (A) FLAIR only, (B) T1 only, (C) FLAIR and T1 concatenated as separate input channels, and (D) FLAIR and T1 treated as interchangeable modalities during training. Configurations (A) and (B) represent unimodal input settings, while (C) and (D) implement alternative strategies for multimodal integration. In configuration (D), the two modalities are considered interchangeable variants, effectively augmenting the training set and encouraging robustness to missing modality scenarios.

The same modeling strategy is applied for both WM lesion and anatomical region segmentation. Specifically, WM region labels are used in place of lesion masks in the architecture shown in Figure~\ref{method_diagram}. Beyond multimodal input strategies, we also investigate joint WM localization, as illustrated in Figure~\ref{method_diagram_2}, where WM lesions and anatomical regions are segmented simultaneously by a shared network. These models are trained with multimodal inputs to predict lesion masks restricted to white matter regions, using masked regional labels (element-wise multiplication of the binary lesion masks with the corresponding WM region labels) as supervision. This design enables the simultaneous estimation of lesion burden and anatomical localization in a single forward pass.

\subsection{White Matter Labels}

Ground truth labels for training the WM region segmentation models were derived from the refined reference labels provided by the JHU MNI White Matter Atlas Type II \cite{oishi2009atlas}. The refined version of the atlas delineates 34 white matter subregions, selected based on ontological hierarchies and clinical relevance. To generate subject-specific label maps, the atlas T1 image was affinely registered to each subject's T1 scan using the extracted WM region \cite{ghazi2022fast}. The resulting transformation was then applied to the atlas region labels, yielding anatomically aligned WM labels in the subject’s native space. The complete preprocessing pipeline is described in detail in \cite{machnio2024deep}.

\subsection{Training and Inference}

All models are based on the 3D U-Net architecture described in \cite{ronneberger2015u}. To improve robustness, we apply extensive MRI-specific data augmentation during training \cite{llambias2023data}. These augmentations include additive and multiplicative noise, bias field distortion, elastic deformations, random rotations, and simulated motion artifacts. Model optimization is performed using a composite loss function that combines cross-entropy (CE) loss with the Dice-Sørensen (DS) loss. During inference, configuration (C) produces a single prediction directly from the concatenated multimodal input. For input configurations (A), (B), and (D), predictions from T1 and FLAIR scans are fused by averaging their softmax outputs, followed by voxel-wise \texttt{ArgMax}.

\section{Experiments and Results} \label{sec:experiments}

\subsection{Data}

We conducted all experiments using the MICCAI 2017 White Matter Hyperintensity (WMH) Segmentation Challenge dataset \cite{kuijf2019standardized}, which contains co-registered 3D FLAIR and T1 MRI scans from 170 subjects across three clinical sites: Utrecht, Amsterdam, and Singapore. The dataset provides expert-annotated lesion masks that differentiate WM lesions from healthy tissue and other pathologies.

To increase the number of samples for training lesion localization models, we inverted the original challenge-defined splits, repurposing the original test set for training. The resulting dataset was then used for 5-fold cross-validation. Summary statistics are provided in Table~\ref{dataset_properties}.

\begin{table}[t]
\centering
\caption{Overview of the WMH data used in this study. The vendor abbreviations refer to GE (G), Philips (P), and Siemens (S).}
\label{tab:dataset_properties}
\resizebox{\columnwidth}{!}{%
\begin{tabular}{lccccc}
\hline
\multicolumn{1}{c}{\textbf{Dataset Split}} & \textbf{\#Subjects}  & \textbf{Dimensions} & \textbf{Resolution} & \textbf{Strength} & \textbf{Vendor} \\ \hline
\textbf{Train}                             & 110                      &181$\times$251$\times$81       &1.1$\times$1$\times$3mm$^3$              & 1.5T,3T           & G, P, S        \\
\textbf{Test}                              & 60                     & 202$\times$250$\times$60      &1.1$\times$0.98$\times$3mm$^3$             & 3T                & G, P, S        \\ \hline
\end{tabular}%
}
\label{dataset_properties}
\end{table}

\subsection{Settings}

For all models, the composite loss function was computed as an equal-weighted sum of CE and DS losses. The 3D U-Net models were implemented in PyTorch \cite{llambias2024yucca} and trained on NVIDIA A100 GPUs with 80~GB of VRAM. Optimization was performed using stochastic gradient descent (SGD) with Nesterov momentum set to 0.9 and an initial learning rate of 0.001. Each model was trained for 1,000 iterations, with up to 250 mini-batches per epoch. In the multi-task setup, we used a batch size of 12 and 3D input patches of size $32 \times 128 \times 128$ voxels.

\subsection{Results}

\paragraph{\textbf{WM Lesion Segmentation}}

We first evaluated the performance of the WM lesion segmentation models on the test set. Table~\ref{tab:lesion} presents the results across different training and inference configurations. The model trained with concatenated T1 and FLAIR inputs achieved the highest Dice score of 0.74, highlighting the advantage of leveraging complementary multimodal information where T1 provides anatomical detail, while FLAIR emphasizes contrast between healthy and pathological tissue.

Models trained on a single modality showed slightly reduced performance: the FLAIR-only model reached a DSC of 0.72, while the T1-only model scored 0.59. The model trained with T1 and FLAIR as interchangeable modalities yielded a lower overall accuracy of 0.67, but it offers unique robustness. This configuration supports inference with just one modality available, making it particularly valuable in clinical contexts where one sequence may be missing or degraded. While slightly less accurate, it provides increased flexibility for real-world deployment.

\begin{table}[t]
\centering
\caption{Test DSC (mean$\pm$SD) for WM lesion segmentation using 3D U-Net models with different training inputs (rows) and inference strategies (columns). The “T1 \& FLAIR” column reports voxel-wise predictions obtained by using softmax outputs from processed T1 and FLAIR inputs.}
\label{tab:lesion}
\resizebox{\columnwidth}{!}{%
\begin{tabular}{lccc}
\toprule
\textbf{Training / Inference} & \textbf{T1} & \textbf{FLAIR} & \textbf{T1 \& FLAIR} \\
\midrule
FLAIR  & - & $0.72 \pm 0.12$ & \multirow{2}{*}{$0.68 \pm 0.14$} \\
T1 & $0.59 \pm 0.16$ & - &  \\
\text{T1 and FLAIR} & - & - & $0.74 \pm 0.11$ \\
\text{T1 or FLAIR} & $0.58 \pm 0.17$ & $0.73 \pm 0.11$ & $0.67 \pm 0.15$ \\
\bottomrule
\end{tabular}%
}
\end{table}

\paragraph{\textbf{WM Region Segmentation}}

We next evaluated the performance of anatomical WM region segmentation. As shown in Table~\ref{tab:regions}, all input configurations achieved similarly high Dice scores, averaging around 0.75, indicating that regional white matter structures can be reliably segmented regardless of modality. Models evaluated using only FLAIR showed a slight reduction in accuracy compared to those using T1 or both modalities, reflecting the lower anatomical contrast inherent in FLAIR images. The multimodal ensemble model produced consistent results across all evaluation settings.

\paragraph{\textbf{WM Lesion Localization}}

Finally, we trained unified models to jointly segment regional WM lesions within a single network. Tables~\ref{tab:together_lesions} and~\ref{tab:together_regions} summarize the corresponding lesion and region segmentation results. Although this approach offers a compact framework for simultaneously predicting regional lesion labels, it exhibited notably reduced performance compared to separate single-task models. In the multimodal configuration, the lesion segmentation Dice score declined from 0.74 to 0.43, while region segmentation dropped from 0.75 to 0.29.

\begin{table}[t]
\centering
\caption{Test DSC (mean$\pm$SD) for WM region segmentation using 3D U-Net models with different training inputs (rows) and inference strategies (columns). The “T1 \& FLAIR” column reports voxel-wise predictions obtained by using softmax outputs from processed T1 and FLAIR inputs.}
\label{tab:regions}
\resizebox{\columnwidth}{!}{%
\begin{tabular}{lccc}
\toprule
\textbf{Training / Inference} & \textbf{T1} & \textbf{FLAIR} & \textbf{T1 \& FLAIR} \\
\midrule
FLAIR  & - & $0.75 \pm 0.05$ & \multirow{2}{*}{$0.75 \pm 0.05$} \\
T1  & $0.75 \pm 0.05$ & - &  \\
\text{T1 and FLAIR} & - & - & $0.75 \pm 0.05$ \\
\text{T1 or FLAIR} & $0.75 \pm 0.05$ & $0.70 \pm 0.06$ & $0.74 \pm 0.06$ \\
\bottomrule
\end{tabular}%
}
\end{table}

\begin{table}[t]
\centering
\caption{Test DSC (mean$\pm$SD) for WM lesion segmentation using 3D U-Nets trained for regional WMH label segmentation with different training inputs (rows) and inference strategies (columns). The “T1 \& FLAIR” column reports voxel-wise predictions obtained by using softmax outputs from T1 and FLAIR inputs.}
\label{tab:together_lesions}
\resizebox{\columnwidth}{!}{%
\begin{tabular}{lccc}
\toprule
\textbf{Training / Inference} & \textbf{T1} & \textbf{FLAIR} & \textbf{T1 \& FLAIR} \\
\midrule
\text{T1 and FLAIR} & - & - & $0.43 \pm 0.20$ \\
\text{T1 or FLAIR} & $0.27 \pm 0.18$ & $0.36 \pm 0.19$ & $0.26 \pm 0.19$ \\
\bottomrule
\end{tabular}%
}
\end{table}

\begin{table}[t]

\centering
\caption{Test DSC (mean$\pm$SD) for WM region segmentation using 3D U-Nets trained for regional WMH label segmentation with different training inputs (rows) and inference strategies (columns). The “T1 \& FLAIR” column reports voxel-wise predictions obtained by using softmax outputs from T1 and FLAIR inputs.}
\label{tab:together_regions}
\resizebox{\columnwidth}{!}{%
\begin{tabular}{lccc}
\toprule
\textbf{Training / Inference} & \textbf{T1} & \textbf{FLAIR} & \textbf{T1 \& FLAIR} \\
\midrule
\text{T1 and FLAIR} & - & - & $0.29 \pm 0.12$ \\
\text{T1 or FLAIR}  & $0.17 \pm 0.11$ & $0.25 \pm 0.11$ & $0.17 \pm 0.11$ \\
\bottomrule
\end{tabular}%
}
\end{table}

\subsection{Discussion}

Our results demonstrate the advantage of multimodal input for WM lesion segmentation, with the highest performance achieved when T1 and FLAIR images were concatenated and jointly processed. This finding reinforces prior work showing that multimodal MRI leverages complementary contrasts: FLAIR enhances lesion visibility due to CSF suppression, while T1 provides clearer anatomical context. Although FLAIR-only models outperformed T1-only models, consistent with FLAIR's superior lesion contrast, the combined input configuration offered improved spatial precision and generalization.

The modality-interchangeable configuration, in which T1 and FLAIR were treated as alternative inputs, yielded lower segmentation performance. Nevertheless, this approach offers a practical advantage: the ability to operate when only one modality is available. This robustness is particularly valuable in real-world clinical scenarios, where incomplete or corrupted data are common. In such settings, the flexibility of this configuration may outweigh the modest reduction in accuracy, especially for large-scale studies or multi-site applications with variable imaging protocols.

For WM region segmentation, performance was more consistent across input types. Most configurations achieved comparable accuracy, with FLAIR-only predictions showing slightly lower performance, likely due to the reduced anatomical contrast in FLAIR images. These results indicate that while FLAIR is well-suited for lesion detection, T1-weighted images remain more informative for anatomical delineation of WM subregions.

In our final set of experiments, we explored a multi-task learning setup where the model jointly segmented lesions and anatomical regions. This configuration resulted in a marked performance drop relative to the single-task models. The reduced accuracy may reflect optimization conflicts or representational interference between the two tasks. To fairly compare the multi-task models with the single-task baselines, we evaluated lesion and region segmentation separately. For lesion assessment, we combined all predicted lesion subregion labels into a single binary mask. For region segmentation, we evaluated predictions only for WM subregions present in each scan. This ensured consistent and representative comparison across all settings.



\section{Conclusion} \label{sec:conclusion}

We presented a systematic study of deep learning strategies for WM lesion segmentation and localization using single- and multimodal MRI inputs. Our framework evaluated multiple input configurations, including unimodal (FLAIR or T1), concatenated multimodal, and modality-interchangeable training. We further extended this setup to jointly segment WM lesions and anatomical subregions via multi-task learning. Experiments on the WMH segmentation dataset demonstrated that combining T1 and FLAIR inputs in a shared model yields the highest segmentation performance, outperforming unimodal baselines. While the modality-interchangeable setup underperformed slightly, it offers robustness in scenarios with missing or incomplete modalities, which is critical for clinical deployment.

Compared to state-of-the-art WMH segmentation approaches, our results reaffirm the importance of multimodal fusion for accurate lesion delineation, and highlight the limitations of single-task models in capturing spatial lesion distribution across anatomical regions. Our multi-task model, designed to jointly segment lesions and WM regions, led to performance degradation, indicating possible interference between task objectives. These findings suggest that while joint learning is promising for efficient inference and spatial lesion quantification, careful architectural and training considerations are necessary. Further investigation is warranted before drawing definitive conclusions. For instance, future work could explore alternative training strategies in which lesion and region labels are treated as separate binary outputs, reducing task entanglement during optimization.

\section*{Acknowledgments}
This project has received funding from the Pioneer Centre for AI, Danish National Research Foundation (DNRF), grant number P1.

{
    \small
    \bibliographystyle{ieeenat_fullname}
    \bibliography{main}
}

\clearpage
\appendix

\end{document}